\documentclass{article}
\usepackage{spconf,amsmath,graphicx}
\usepackage{algorithm}
\usepackage{algorithmic}
\usepackage{amssymb}
\usepackage{multirow}
\usepackage{url}
\usepackage{tabto}

\title{Federated Learning for Keyword spotting}
\name{David Leroy, Alice Coucke, Thibaut Lavril, Thibault Gisselbrecht and Joseph Dureau}
\address{Snips, 18 rue Saint Marc, 75002 Paris, France}
\begin{document}
%
\maketitle
\begin{abstract}

\begin{sloppypar}
We propose a practical approach based on federated learning to solve out-of-domain issues with continuously running embedded speech-based models such as wake word detectors. We conduct an extensive empirical study of the federated averaging algorithm for the ``Hey Snips'' wake word based on a crowdsourced dataset that mimics a federation of wake word users.  We empirically demonstrate that using an adaptive averaging strategy inspired from Adam in place of standard weighted model averaging highly reduces the number of communication rounds required to reach our target performance. The associated upstream communication costs per user are estimated at 8 MB, which is a reasonable in the context of smart home voice assistants. Additionally, the dataset used for these experiments is being open sourced with the aim of fostering further transparent research in the application of federated learning to speech data.

\end{sloppypar}

\end{abstract}


\vspace{-2mm}
\section{Introduction}
\label{sec:intro}
\vspace{-2mm}

\begin{sloppypar}
Wake word detection is used to start an interaction with a voice assistant. A specific case of keyword spotting (KWS), it continuously listens to an audio stream to detect a predefined keyword or set of keywords. Well-known examples of wake words include Apple's ``Hey Siri''  or Google's ``OK Google''. Once the wake word is detected, voice input is activated and processed by a spoken language understanding engine, powering the perception abilities of the voice assistant \cite{coucke2018svp}. 

Wake word detectors usually run on device in an always-on fashion, which brings two major difficulties. First, it should run with minimal memory footprint and computational cost. The resource constraints for our wake word detector are 200k parameters (based on the medium-sized model proposed in \cite{zhang2017kws}), and 20 MFLOPS.

Secondly, the wake word detector should behave consistently in any usage setting, and show robustness to background noise. The audio signal is highly sensitive to recording proximity (close or far field), recording hardware, but also to the room configuration. Robustness also implies a strong speaker variability coverage. While the use of digital signal processing front-ends can help mitigate issues related to bad recording conditions, speaker variability remains a major challenge. High accuracy is all the more important since the model can be triggered at any time: it is therefore expected to capture most of the commands (high recall) while not triggering unintentionally (low false alarm rate). 
\end{sloppypar}

\begin{sloppypar}
Today, wake word detectors are typically trained on datasets collected in real usage setting e.g. users homes in the case of voice assistants. Speech data being by nature very sensitive, centralized collection raises major privacy concerns. In this work, we investigate the use of federated learning (FL) \cite{mcmahan2016fl} in the context of an embedded wake word detector. FL is a decentralized optimization procedure that enables to train a central model on the local data of many users without the need to ever upload this data to a central server. The training workload is moved towards the user's devices which perform training steps on the local data. Local updates from users are then averaged by a parameter server in order to create a global model.
\end{sloppypar}

\vspace{-2mm}
\section{Related work}
\label{sec:related_work}
\vspace{-2mm}

\begin{sloppypar}

Most research around decentralized learning has historically been done in the context of a highly controlled cluster/data center setting, e.g. with a dataset evenly partitioned in an i.i.d fashion. The multi-core and multi-gpu distributed training setting has been specifically studied in the context of speech recognition in \cite{povey15disttrain}. Efforts on decentralized training with highly distributed, unbalanced and non-i.i.d data is relatively recent, as the foundations were laid down in \cite{mcmahan2016fl} with the introduction of the federated averaging (\textit{FedAvg}) algorithm and its application to a set of computer vision (MNIST, CIFAR-10) and language modeling tasks (applied to the Shakespeare and Google Plus posts datasets).
To our knowledge, the present work is the first experiment of its kind on user-specific speech data.

The federated optimization problem in the context of convex objective functions has been studied in \cite{konecny2016fo}. The authors proposed a stochastic variance-reduced gradient descent optimization procedure (\textit{SVRG}) with both local and global per-coordinate gradient scaling to improve convergence. Their global per-coordinate gradient averaging strategy relies on a sparsity measure of the given coordinate in users local datasets and is only applicable in the context of sparse linear-in-the-features models. The latter assumption does not hold in the context of neural networks for speech-based applications. 

Several improvements to the initial \textit{FedAvg} algorithm have been suggested with a focus on client selection \cite{nishio2018fedcs}, budget-constrained optimization \cite{wang2018adptativefl} and upload cost reduction for clients \cite{konecny2016fl}. A dynamic model averaging strategy robust to concept drift based on a local model divergence criterion was recently introduced in \cite{kamp2018dynmodelavg}. While these contributions present efficient strategies to reduce the communication costs inherent to federated optimization, the present work is as far as we know the first one introducing a dynamic per-coordinate gradient update in place of the global averaging step.
\end{sloppypar}

\begin{sloppypar}
The next section describes the federated optimization procedure, and how its global averaging can be substituted by an adaptative averaging rule inspired from Adam. It is followed by the experiments section, where both the open-sourced crowdsourced data and model used to train our wake word detector are introduced. Results come next, and a communication cost analysis is provided. Finally, the next steps towards training a wake word detector on user data that is really decentralized are described.
\end{sloppypar}

\vspace{-2mm}
\section{Federated optimization}
\label{sec:fo}
\vspace{-2mm}

We consider the standard supervised learning objective function $f_{i}(w) = l(x_i,y_i,w)$ that is the loss function for the prediction on example $(x_i, y_i)$ when using a model described by a real-valued parameter vector $w$ of dimension $d$. In a federated setting, we assume that the datapoints $i$ are partitioned across $K$ users, each user being assigned their own partition $\mathcal{P}_{k}$, $|\mathcal{P}_{k}|=n_k$. The optimization objective is therefore the following:

\begin{equation}
\begin{array}{c}
\underset{w \in \mathbb{R}^{d}}{\mathrm{min}} f(w) \ \ \mathrm{where} \ \ f(w) \overset{\mathrm{def}}{=}\sum_{k=1}^{K}\frac{n_k}{n}\times F_k(w) \\ \\
\mathrm{with} \ \ F_{k}(w) = \frac{1}{n_k}\sum_{i=1}^{n_k}f_i(w)
\end{array}
\label{equ:opt_objective}
\end{equation}

The \textit{FedAvg} algorithm introduced in \cite{mcmahan2016fl} aims at minimizing the objective function \ref{equ:opt_objective} assuming a synchronous update scheme and a generic non-convex neural network loss function. The model is initialized with a given architecture on a central \textit{parameter server} with weights $w_{0}$. Once initialized, the parameter server and the user's devices interact synchronously with each other during \textit{communication rounds}. A communication round at time $t \in [1, ..,T]$ is described below:
\begin{enumerate}
    \vspace{-2mm}\item The central model $w_{t-1}$ is shared with a subset of users $\mathcal{S}_{t}$ that are randomly selected from the pool of \textit{K} users given a participation ratio $\textit{C}$.
    \vspace{-2mm}\item Each user $k \in \mathcal{S}_{t}$ performs one or several training steps on their local data based on the minimization of their local objective $F_{k}$ using mini-batch stochastic gradient descent (SGD) with a local learning rate $\eta_{local}$. The number of steps performed locally is \hfill\\ $E\times max(ceil(\frac{n_k}{B}),1)$, $n_k$ being the number of datapoints available locally, \textit{E} the number of local epochs and \textit{B} the local batch size.
    \vspace{-2mm}\item Users from $\mathcal{S}_{t}$ send back their model updates $w_{t,k}, k\in \mathcal{S}_{t}$ to the parameter server once local training is finished. 
    \vspace{-2mm}\item The server computes an average model $w_{t}$  based on the user's individual updates $w_{t,k} , k\in \mathcal{S}_{t}$, each user's update being weighted by $\frac{n_k}{n_r}$, ‘’’where’’’  $n_r=\sum_{k \in \mathcal{S}_{t}}n_k \approx  C\times \sum_{k=1}^{K}n_{k}$. 
\end{enumerate}

When $B=\infty$ (i.e the batch size is equal to the local dataset size) and $E=1$, then a single gradient update is performed on each user's data. It is strictly equivalent to doing a single gradient computation on a batch including all of selected user data points. This specific case is called \textit{FedSGD}, e.g. stochastic gradient descent with each batch being the data of the federation of selected users at a given round. \textit{FedAvg} (Federated averaging) is the generic case when more than one update is performed locally for each user. 

The global averaging step can be written as follows, using a global update rate $\eta_{global}$.

\begin{equation}
w_{t} \leftarrow w_{t-1} - \eta_{global}\sum_{k \in \mathcal{S}_{t}}\frac{n_k}{n}( w_{t-1} - w_{t,k} )
\label{equ:global_optimization}
\end{equation}

Setting the global update rate $\eta_{global}$ to 1 is equivalent to a weighted averaging case without moving average. Equation \ref{equ:global_optimization} highlights the parallel between global averaging and a gradient update $\mathcal{G}_{t} = \sum_{k \in \mathcal{S}_{t}}\frac{n_k}{n}( w_{t-1} - w_{t,k} )$. This parallel motivates the use of adaptive per-coordinate updates for $\mathcal{G}_{t}$ that have proven successful for centralized deep neural networks optimization such as Adam \cite{kingma2014adam}. Moment-based averaging allows to smooth the averaged model by taking into account the previous rounds updates that were computed on different user subsets. We conjecture that the exponentially-decayed first and second order moments perform the same kind of regularization that occurs in the mini-batch gradient descent setting, where Adam has proven to be successful on a wide range of tasks with various neural-network based architectures. In this work, we set the exponential decay rates for the moment estimates to $\beta_{1}=0.9$ and $\beta_{2}=0.999$ and $\epsilon = 10^{-8}$ as initially suggested by the authors of \cite{kingma2014adam}. 

In the federated setting, model evaluation is also done in a distributed fashion on a set of users that are kept specifically for this purpose. Model metrics are averaged on the \textit{parameter server} with a similar weighting scheme as parameter averaging.

\vspace{-2mm}
\section{Experiments}
\label{sec:experiments}
\subsection{Dataset}
\vspace{-2mm}

Unlike generic speech recognition tasks, there is no reference dataset for wake word detection. The reference dataset for multi-class keyword spotting is the speech command dataset \cite{warden18sc}, but the speech command task is generally preceded by a wake word detector and is focused on minimizing the confusion across classes, not robustness to false alarms. We constituted a crowdsourced dataset for the \textit{Hey Snips} wake word. We are releasing publicly \footnote{\url{http://research.snips.ai/datasets/keyword-spotting}} this dataset \cite{coucke2018efficient} in the hope it can be useful to the keyword spotting community.

\begin{table}[ht]
    \centering
      \begin{tabular}{l|l|l|l}
        \hline
        Train set & Dev set & Test set & Total \\
        \hline
        1,374 users & 200 users & 200 users & 1774 users \\
        53,991 utt. & 8,337 utt. & 7,854 utt. & 69,582 utt. \\
        \hline
    \end{tabular}
    \caption{Dataset statistics for the \textit{Hey Snips} wake word - 18\% of utterances are positive, with strong per user imbalance in the number of utterances \textit{(mean: 39, standard dev: 32)}}
    \label{tab:dataset_stats}
\end{table}

The data used here was collected from 1.8k contributors that recorded themselves on their device with their own microphone while saying several occurrences of the \textit{Hey Snips} wake word along with negative short sentences from various text sources (e.g. subtitles).


This crowdsourcing-induced data distribution mimicks a real-world non-i.i.d, unbalanced and highly distributed setting, and a parallel is therefore drawn in the following work between a crowdsourcing contributor and a voice assistant user. The statistics about the dataset comforting this analogy are summarized in Table \ref{tab:dataset_stats}. The train, dev and test splits are built purposely using distinct users, 77\% of users being used solely for training while the remaining are used for parameter tuning and final evaluation, measuring the generalization power of the model to new users.

\vspace{-2mm}
\subsection{Model}
\vspace{-2mm}

Acoustic features are generated based on 40-dimensional mel-frequency cepstrum coefficients (MFCC) computed every 10ms over a window of 25ms. The input window consists in 32 stacked frames, symmetrically distributed in left and right contexts. The architecture is a CNN with 5 stacked dilated convolutional layers of increasing dilation rate, followed by two fully-connected layers and a softmax inspired from \cite{peddinti2015time}. The total number of parameters is 190,852, initialized using Xavier initialization \cite{Glorot10understandingthe}. The model is trained using cross-entropy loss on frames prediction. The neural network has 4 output labels, assigned via a custom aligner specialized on the target utterance ``Hey Snips'': ``Hey'', ``sni'', ``ps'', and ``filler'' that accounts for all other cases (silence, noise and other words). A posterior handling \cite{chen2014small} generates a confidence score for every frame by combining the smoothed label posteriors. The model triggers if the confidence score reaches a certain threshold $\tau$, defining the operating point that maximizes recall for a certain amount of False Alarms per Hour (FAH). We set the number of false alarms per hour to 5 as a stopping criterion on the dev set. The dev set is a ``hard'' dataset when it comes to false alarms since it belongs to the same domain as data used for training. The model recall is finally evaluated on the test set positive data, while false alarms are computed on both the test set negative data and various background negative audio sets. See section \ref{subsec:results} for further details about evaluation.

\vspace{-2mm}
\subsection{Results}
\label{subsec:results}
\vspace{-2mm}

We conduct an extensive empirical study of the federated averaging algorithm for the \textit{Hey Snips} wake word based on crowdsourced data from Table \ref{tab:dataset_stats}. Federated optimization results are compared with a \textit{standard setting} e.g. centralized mini-batch SGD with data from train set users being randomly shuffled. Our aim is to evaluate the number of communication rounds that are required in order to reach our stopping criterion on the dev set. For the purpose of this experiment early stopping is evaluated in a centralized fashion, and we assume that the dev set users agreed to share their data with the parameter server. In an actual product setting, early stopping estimation would be run locally on the devices of the dev users, they would download the latest version of the central model at the end of each round and evaluate the early stopping criterion based on prediction scores for their own utterances. These individual metrics would then be averaged by the parameter server to obtain the global model criterion estimation. Final evaluation on test users would be done in the same distributed fashion once training is finished.

\textbf{Standard baseline}: Our baseline e.g. a standard centralized data setting with a single training server and the Adam optimizer reaches the early stopping target in 400 steps ($\sim2$ epochs) and is considered as our upper bound of performance. Adam provides a strong convergence speedup in comparison with standard SGD that remains under 87\% after 28 epochs despite learning rate and gradient clipping tuning on the dev set.

\begin{figure}[!ht]
    \centering
    \includegraphics[width=8.5cm]{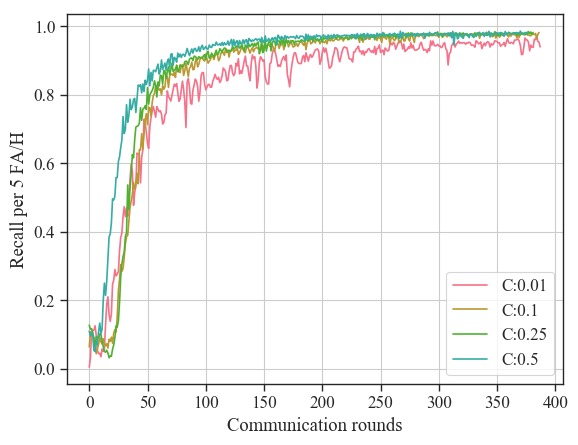}
    \caption{Effect of the share of users involved in each round \textit{C} on the dev set recall / 5 FAH, \textit{FedSGD}, Adam global averaging, $\eta_{global}= 0.001, \eta_{local} = 0.01$}
    \label{fig:c_search}
\end{figure}

\textbf{User parallelism}: The higher the ratio of users selected at each round \textit{C}, the more data is used for distributed local training, and the faster the convergence is expected, assuming that local training does not diverge too much. Figure \ref{fig:c_search} shows the impact of \textit{C} on convergence: the gain of using half of users is limited with comparison with using 10\%, specifically in the later stages of convergence. A fraction of 10\% of users per round is also more realistic in a practical setup as selected users have to be online. With lower participation ratios ($C=1\%$), the gradients are much more sensitive and might require the use of learning rate smoothing strategy. \textit{C} is therefore set to 10\%.

\textbf{Global averaging}: Global adaptive learning rates based on Adam accelerates convergence when compared with standard averaging strategies with or without moving averages. Table \ref{tab:results_ga} summarizes experimental results in the \textit{FedSGD} setting with optimized local learning rates. Applying standard global averaging yields poor performances even after 400 communication rounds when compared with adaptive per-parameter averaging.

\begin{table}[!ht]
    \centering
    \begin{tabular}{l|c|c}
        \hline
         Avg. Strategy & 100 rounds  & 400 rounds \\
        \hline
        $\eta_{global}=1.0$ & 29.9\% & 67.3\%\\
        \hline
        Adam & & \\
        $\eta_{global}=0.001$ & 93.50\%& 98.29\% \\
        \hline
    \end{tabular}
    \caption{
        Dev set recall / 5 FAH for various averaging strategies - \textit{FedSGD},  $\textit{C}=10\%$
    }
    \label{tab:results_ga}
\end{table}

\textbf{Local training}: 
Our results show consistency across local training configurations, with limited improvements coming from increasing the load of local training. The number of communication rounds required to reach the stopping criterion on the dev set ranges between 63 and 112 communication rounds for $\textit{E} \in [1,3]$ and $\textit{B}\in [20, 50, \infty]$, using $C=10\%$, \textit{Adam} global averaging with $\eta_{global}= 0.001$, and a local learning rate of 0.01. In our experiments, the best performances are obtained for $E=1$ and $B=20$ for an average of 2.4 local updates per worker taking part in a round, yielding a 80\% speedup with comparison to \textit{FedSGD}. The total number of training steps needed to reach the stopping criterion amounts to approximately $3300$ for $100$ communication rounds e.g. $8.25$ times the number of steps required in the standard setting, with much smaller batches.  Nevertheless we observed variability across experiments with regard to random weight initialization and early stage behaviour. Unlike some experiments presented in \cite{mcmahan2016fl}, the speedup coming from increasing the amount of local training steps does not lead to order of magnitude improvements on convergence speed, while local learning rate and global averaging tuning proved to be crucial in our work. We conjecture that this difference is related to the input semantic variability across users. In the MNIST and CIFAR experiments from \cite{mcmahan2016fl} the input semantics are the same across emulated users. In the wake word setting, each user has their own vocalization of the same wake word utterance with significant differences in pitch and accent than can lead to diverging lower stage representations that might perform poorly when averaged.

\textbf{Evaluation}: We evaluate the false alarm rates of the best model ($E=1$ and $B=20$) for a fixed recall of 95\% on the test set. We observe 3.2 FAH on the negative test data, 3.9 FAH on Librispeech \cite{panayotovCPK15librispeech}, and respectively 0.2 and 0.6 FAH on our internal news and collected TV datasets. Unsurprisingly, false alarms are more common on close-field continuous datasets than they are on background negative audio sets.


\vspace{-2mm}
\subsection{Communication cost analysis}
\vspace{-2mm}
Communication cost is a strong constraint when learning from decentralized data, especially when user’s devices have limited connectivity and bandwidth. Considering the asymmetrical nature of broadband speeds, the communication bottleneck for federated learning is the updated weights transfer from clients to the parameter server \cite{konecny2016fl}. We assume that the upstream communication cost associated with users involved in model evaluation at each communication round is marginal, as they would only be uploading a few floating point metrics per round that is much smaller than the model size. The total client upload bandwidth requirement is provided in the equation below:
\begin{equation}
\rm{ClientUploadCost} = \rm{modelSize}_{f32}\times C \times N_{rounds}
\label{equ:com_cost}
\end{equation}

Based on our results, this would yield a cost of 8MB per client when the stopping criterion is reached within 100 communication rounds. The server receives 137 updates per round when $\textit{C}=10\%$, amounting for 110GB over the course of the whole optimization process with 1.4k users involved during training. Further experiments with latter convergence stages (400 rounds) yielded 98\% recall / 0.5 FAH on the test set for an upload budget of 32 MB per user.

\vspace{-2mm}
\section{Conclusion and future Work}
\label{sec:conclusion}
\vspace{-2mm}

In this work, we investigate the use of federated learning on crowdsourced speech data to learn a resource-constrained wake word detector. We show that a revisited \textit{Federated Averaging} algorithm with per-coordinate averaging based on Adam in place of standard global averaging allows the training to reach a target stopping criterion of 95\% recall per 5 FAH within 100 communication rounds on our crowdsourced dataset for an associated upstream communication costs per client of 8MB. We also open source the \textit{Hey Snips} wake word dataset.

The next step towards a real-life implementation is to design a system for local data collection and labeling as the wake word task requires data supervision. The frame labeling strategy used in this work relies on an aligner, which cannot be easily embedded. The use of of memory-efficient end-to-end models \cite{coucke2018efficient} in place of the presented class-based model could ease the work of local data labelling. 

\newpage

\bibliographystyle{IEEE}

\end{document}